\documentclass[aps,pre,amsmath,amssymb,12pt,tightenlines]{revtex4}

\usepackage{graphicx}
\usepackage{amsmath,amssymb,bm,amsthm}
\begin{document}
\title{Visualization by optical fluorescence of two-phase flow in a three-dimensional porous medium}

\author{Joachim Falck Brodin$^1$, Marcel Moura$^1$, Renaud Toussaint$^2$,\\ 
Knut J{\o}rgen M{\aa}l{\o}y$^1$ and Per Arne Rikvold$^{1,3~}$}
\email{j.f.brodin@fys.uio.no, marcel.moura@fys.uio.no, 
renaud.toussaint@unistra.fr, k.j.maloy@fys.uio.no, prikvold@fsu.edu}

\affiliation{$^1$ PoreLab, NJORD Centre, Department of Physics, University of Oslo, 
P.O.\ Box 1048 Blindern, 0316 Oslo, Norway\\
$^2$ Universit{\'e} de Strasbourg, CNRS, Institut de Physique du Globe de Strasbourg,
UMR 7516, France\\
$^3$ Department of Physics, Florida State University, Tallahassee, FL 32306-4350, USA}

\begin{abstract}
Slow flow of a single fluid through a porous medium is well understood on a macroscopic level 
through Darcy’s law, a linear relation between flow rate and a combination of 
pressure differences, viscosity, and gravitational forces. 
Two-phase flow is complicated by 
the interface separating the fluids, but understanding of   
two-dimensional, two-phase flow has been obtained from experiments using 
transparent cells. In most 
three-dimensional media, however, visual observation is difficult. Here, we present preliminary results of experiments on a model medium consisting of randomly packed glass spheres, in which one fluorescent liquid invades another. By refractive index matching and scanning with a 
sheet-shaped laser beam, we obtain slices of the flow patterns, which we combine into three-dimensional pictures. We observe a compact region of invading fluid, surrounded by finger-like protrusions. The compact region becomes more dominant with increasing invader flow rate. 
The patterns are theoretically analyzed 
in terms of the interplay between gravitational, viscous, and capillary forces. 
\end{abstract}

\maketitle

\section{Introduction}

Many systems, 
from host rocks in aquifers and oil fields to diapers, kitchen sponges, and biological tissues, 
can be described as porous materials. 
The behavior of fluids in porous media has been under quantitative study at least since the 
19th century, when the French water-works engineer, Henry Darcy, studied the flow of water
through sand-filled pipes \cite{DARCY1856}. 
Macroscopically, slow (``creeping") flow of a single fluid through a porous medium is well 
described by his results. They are summarized by ``Darcy’s Law," 
a linear-response relation between fluid flow rate, pressure differences, viscosity, and gravity. 
Applied to a vertical pipe of constant cross-sectional area $A$, into which fluid enters 
with a constant, downward flow rate $Q$ at $z=0$, it can be written in modern notation as 
\begin{equation}
\label{eq:Darcy}
p(z)  = p(0) -   \left( \rho g - \frac{\mu}{k} \frac{Q}{A} \right) z ~.
\end{equation}
Here, $p(z)$ is the pressure at $z$, $\rho$ is the fluid density, $g$ is the 
acceleration of gravity, $\mu$ is the fluid viscosity, and $k$ is known as the 
{\it permeability}. The latter is a characteristic of the 
medium, which can be considered as an 
effective cross-sectional area of the channels through the microscopic pore structure. 
The ratio $Q/A$ is known as the {\it filtration velocity}. 

Two-phase flow is complicated by the capillary forces at the interface separating the fluids. Understanding of the patterns arising from 
competing gravitational, viscous, and capillary forces for two-dimensional (2D) two-phase flow \cite{BIRO95,Meheust2002a,TOUS12,VASS13} has been obtained using transparent plates
 confining a thin layer of the porous 
medium, enabling visual observation of the flow. Such observation is difficult in 
three dimensions (3D). 

In Sec.\ \ref{sec:Exp}, we present preliminary results from our experiments on a 
3D model medium consisting of randomly packed glass spheres, in which one fluorescent fluid is invaded by another. 
In Sec.\ \ref{sec:Anal}, the observations are analyzed and 
semi-quantitatively explained, using an approximate 
generalization of Eq.~(\ref{eq:Darcy}) that  describes the effects of competing forces in the 
3D flow, analogous to previous 2D results \cite{BIRO95,Meheust2002a,TOUS12,VASS13}. 

\begin{figure}[t]
\begin{center}
\includegraphics[width=0.99\textwidth]{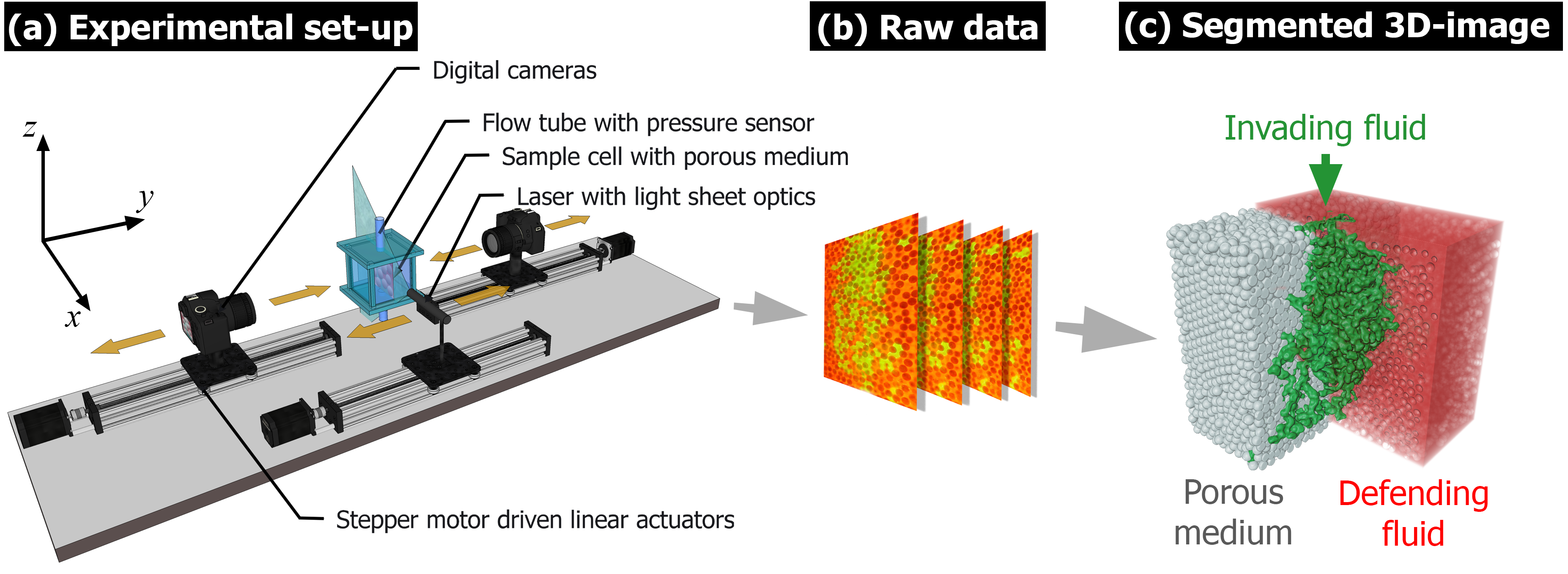}
\end{center}
\caption{{\bf(a):} The experimental set-up to produce 3D images of two-phase flow in 
a porous medium by laser-induced fluorescence. 
The box between the 
cameras is the flow cell, containing a random packing of glass spheres. 
Initially, the medium is saturated with 
rapeseed oil tagged for red fluorescence, which is later invaded from above by glycerol tagged for 
green fluorescence. 
{\bf (b):} Some of the 2D slice images making up the raw data. 
{\bf (c):} Segmented 3D cut-away image at the moment the glycerol percolates to the bottom. 
The gray section shows the porous medium, the red section shows 
the oil, and the green object is the invading glycerol body. 
}\label{fig:1}
\end{figure}

\section{Experimental set-up and results}
\label{sec:Exp}

We consider a situation, in which a dense, viscous liquid  
(glycerol, $\rho_1 = 1.255$~g/ml, $\mu_1 = 1.350$~Pa\,s) is injected from above 
into a porous medium, which is initially saturated with a less dense and much less viscous liquid 
(rapeseed oil,  $\rho_2 = 0.912$~g/ml, $\mu_2 = 0.058$~Pa\,s). 

The experimental set-up is illustrated in Fig.~\ref{fig:1}. 
An $80 \times 80 \times 80$~mm glass box with inlet and outlet pipes 
(radius $r_{\rm in} = 1.5$ mm) 
at top and bottom, respectively, is filled with a random packing of 
glass spheres (radius $r_0= 1.5$ mm and measured porosity $\phi = 0.41$). 
Initially, the pore space between 
the spheres is saturated with the oil tagged with a red, fluorescent dye. Then, glycerol (tagged for 
green fluorescence) is injected through the inlet at 
controlled, constant flow rates between 0.1 and 1.0 ml/min, until the invading 
glycerol reaches the bottom (percolation). 

The 3D flow patterns are visualized by scanning a vertical, sheet-shaped laser beam across 
the sample at 3.5 mm/s, recording 2D ``slice" images of the fluorescence with digital cameras at 
a rate of approximately 49 frames/s. 
Optical transparency is ensured by matching the refractive indices of the 
fluids and the medium \cite{STOH03,DALB18}. 
The cameras are moved horizontally, so as to maintain constant 
optical distance between each camera and the light sheet. 3D images and movies are constructed from the 2D slices, using the Amira-Avizo software from Thermo Scientific. 

\begin{figure}[t]
\begin{center}
\includegraphics[width=1.0\textwidth]{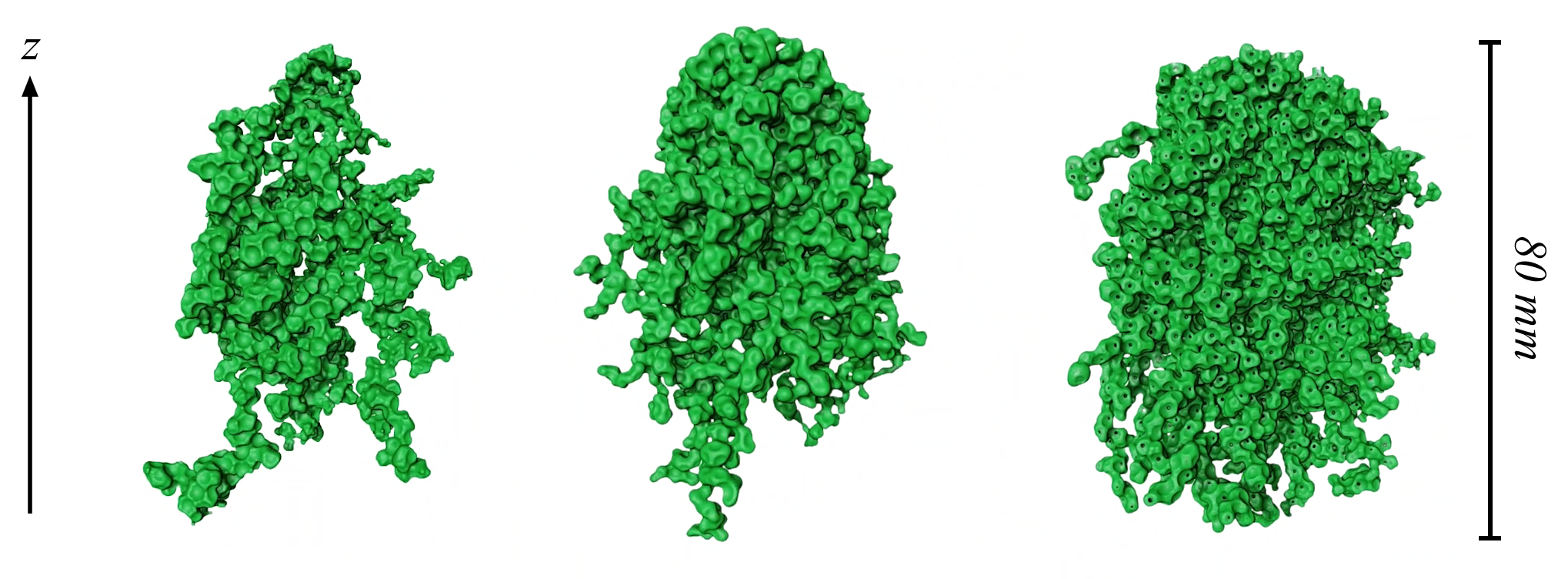}
\end{center}
\caption{Images of the 3D configuration of the invading glycerol liquid at percolation. 
Flow rates, from left to right: $Q = 0.1$, 0.5, and 1.0~ml/min.  
The structures consist of a compact region, surrounded by thin ``fingers." In these pictures, the 
latter are best seen in the lower part and at the sides of the body of glycerol. 
The compact centers  grow with increasing $Q$, and also with time. 
}\label{fig:blob}
\end{figure}

Images of the invading glycerol  
configuration at percolation, produced at different flow rates $Q$, are shown 
in Fig.~\ref{fig:blob}. We observe that the configurations are more compact near the center, 
while extending long ``fingers" near the periphery. The compact region 
grows as $Q$ is increased. 

\section{Theoretical analysis}
\label{sec:Anal}

To estimate the distance below the inlet, where the interface becomes unstable with 
respect to gravitational fingering, we introduce a simple, 
3D generalization of Darcy's law, Eq.~(\ref{eq:Darcy}). In this picture, we approximate the 
filtration velocity of the invading fluid as uniform over a hemispherical surface, 
centered at the inlet. For simplicity, we ignore 
boundary conditions at the outlet, walls, and floor of the box. 

The pressure difference between the two fluids at their interface is known as the 
{\it capillary pressure difference}. 
Its value across the interface at a point a distance $R$ from the inlet and at a depth $-z >0$, 
$(R,z)$, is estimated by integrating the Darcy equation along two paths from the inlet to $(R,z)$, 
each lying entirely inside one of the fluids. 
\begin{equation}
\Delta p(R,z,t)  
\approx  
\Delta p(0,0,t)   - z g \Delta \rho 
- \frac{\mu_1 Q}{ 2 \pi k} \int_{r_{\rm in}}^{R} \frac{{\rm d} r}{r^2}
=\Delta p(0,0,t) - z g \Delta \rho 
+ \frac{\mu_1 Q}{ 2 \pi k} \left[ \frac{1}{R} - \frac{1}{r_{\rm in}} \right] ~,
\label{eq:2}
\end{equation}
where $\Delta \rho = \rho_1 - \rho_2$,  
and $0 \le r_0 \le -z \le R$. 
The permeability for this medium is estimated from the Kozeny-Carman relation 
\cite{DALB18} as 
$k \approx \frac{r_0^2}{45}\frac{\phi^3}{(1-\phi)^2} = 0.99\times 10^{-8}\, {\rm m}^2$.
$\Delta p(0,0,t)$ is the time-dependent pressure applied by the pump at the inlet to maintain 
the flow rate constant at $Q$. 
\begin{figure}[t]
\begin{center}
\includegraphics[width=0.49\textwidth]{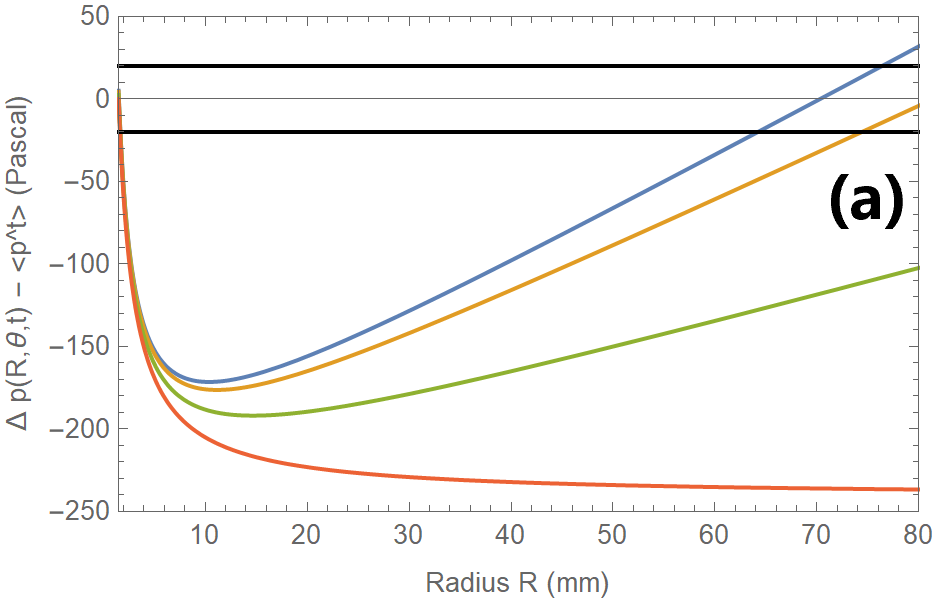}
\includegraphics[width=0.49\textwidth]{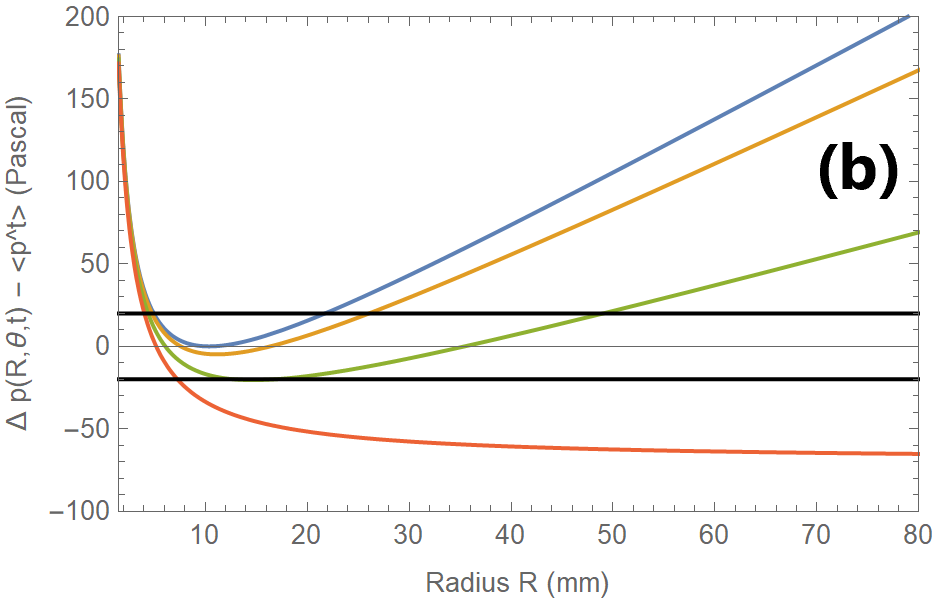}
\end{center}
\caption{The capillary-pressure function $\Delta p(R,\theta,t) - \langle p^{\rm t} \rangle$ 
for $Q = 1.0$~ml/min, vs distance $R$ 
from the inlet, along straight lines at angles $\theta$ with the negative $z$ axis. From 
top to bottom are $\theta = 0^\circ$ (vertically down), $30^\circ$, $60^\circ$, and 
$90^\circ$ (horizontal).  Parts (a) and (b) represent different times, and the 
horizontal, black lines represent the capillary-pressure interval, 
$[\langle p^{\rm t} \rangle -W^{\rm t},\, \langle p^{\rm t} \rangle +W^{\rm t}]$.
{\bf (a):} At a very early time, when $\Delta p(0,t)- \langle p^{\rm t} \rangle \approx 0$. 
The invading glycerol can then only penetrate a very short distance through the largest pores. 
To drive the flow further at constant $Q$, the pump increases $\Delta p(0,t)$. 
{\bf (b):} At a later time,  
when the flow is just becoming unstable toward gravitational fingering near the saddle point at 
the minimum for $\theta = 0^\circ$. 
}\label{fig:3}
\end{figure}

For the interface to progress past a point $\vec{R}$ at $(R,z)$,
the capillary pressure difference at that point, 
$\Delta p(R,z,t) = p_1(R,z) - p_2(R,z)$, must 
exceed a local pressure threshold, $p^{\rm t}(\vec{R})$. Each threshold represents the 
capillary pressure required for the meniscus (surface tension 
$\sigma \approx 16.23 \times 10^{-3}$~N/m) 
to pass from one pore to a neighboring one through a narrow {\it pore throat} at this location.
We assume that these thresholds are 
independent random variables with a probability density of mean $\langle p^{\rm t} \rangle$ 
and a width $W^{\rm t}$ of comparable magnitude. 
Using the Young-Laplace equation, we approximate 
$W^{\rm t} \approx \langle p^{\rm t} \rangle \approx {2 \sigma}/{r_0} 
\approx 20 \, {\rm Pa}$.
For simplicity, we use spherical coordinates to write $-z = R \cos \theta >0$, 
were $\theta$ is the angle between a straight line from the inlet and the negative $z$ axis,
and rename quantities in  Eq.~(\ref{eq:2}) to write the threshold requirement as 
\begin{equation}
\Delta p(R,\theta,t) 
\approx \Delta p(0,0,t) +G R \cos \theta +M \left[ \frac{1}{R} - \frac{1}{r_{\rm in}} \right] 
> p^{\rm t}(\vec{R}) \;.
\label{eq:3}
\end{equation} 
With experimental values from Sec.~\ref{sec:Exp}, $G= 3.36$~Pa/mm and 
$M = 362$~Pa\,mm for $Q = 1.0$~ml/min. 

In Fig.\ \ref{fig:3} we plot,  at two different times, 
$\Delta p(R,\theta,t) - \langle p^{\rm t} \rangle$ 
vs $R$ for different angles $\theta$. 
At both times, the two thick, black lines mark a pressure-threshold interval of width 
$2 W^{\rm t}$, centered at $\langle p^{\rm t} \rangle$. 
Part (a) shows an early time, when $\Delta p(0,0,t) \approx \langle p^{\rm t} \rangle$, 
and part (b) shows a later 
time, when the flow is just becoming unstable against gravitational fingering, as discussed below.

The spatially constant terms in Eq.~(\ref{eq:3}) vanish in the 
gradient of $\Delta p(R,\theta,t)$, which in spherical coordinates is,
\begin{equation}
\nabla [  \Delta p(R,\theta,t) ] 
\approx \hat{r} \left(G \cos \theta - \frac{M}{R^2} \right) - \hat{\theta} G \sin \theta \;.
\label{eq:4}
\end{equation} 
The gravitational driving term, $G \cos \theta$,  and the viscous drag term, $-M/R^2$, 
compete in the radial component (proportional to what is known as the ``modified Bond number" \cite{Meheust2002a}). The latter dominates near the inlet (negative slopes in 
Fig.\ \ref{fig:3}), while the former dominates far below the inlet (positive slopes). 
The gradient vanishes at a point on the negative $z$ axis, with  
\begin{equation}
R_0 = -z = \sqrt{M/G} \;.
\label{eq:R0}
\end{equation}
This is a saddle point for the capillary pressure difference. 
For the compact cluster of invading fluid, $R_0$ is a critical size, beyond which its 
interface against the defending fluid becomes unstable toward gravitationally driven finger growth. 
From Eq.~(\ref{eq:R0}), our estimates for $R_0$ are  approximately 10.4~mm 
for $Q = 1.0$~ml/min, 7.3~mm for $0.5$~ml/min, and 3.3~mm for $0.1$~ml/min. 

\begin{figure}[t]
\begin{center}
\includegraphics[width=1.0\textwidth]{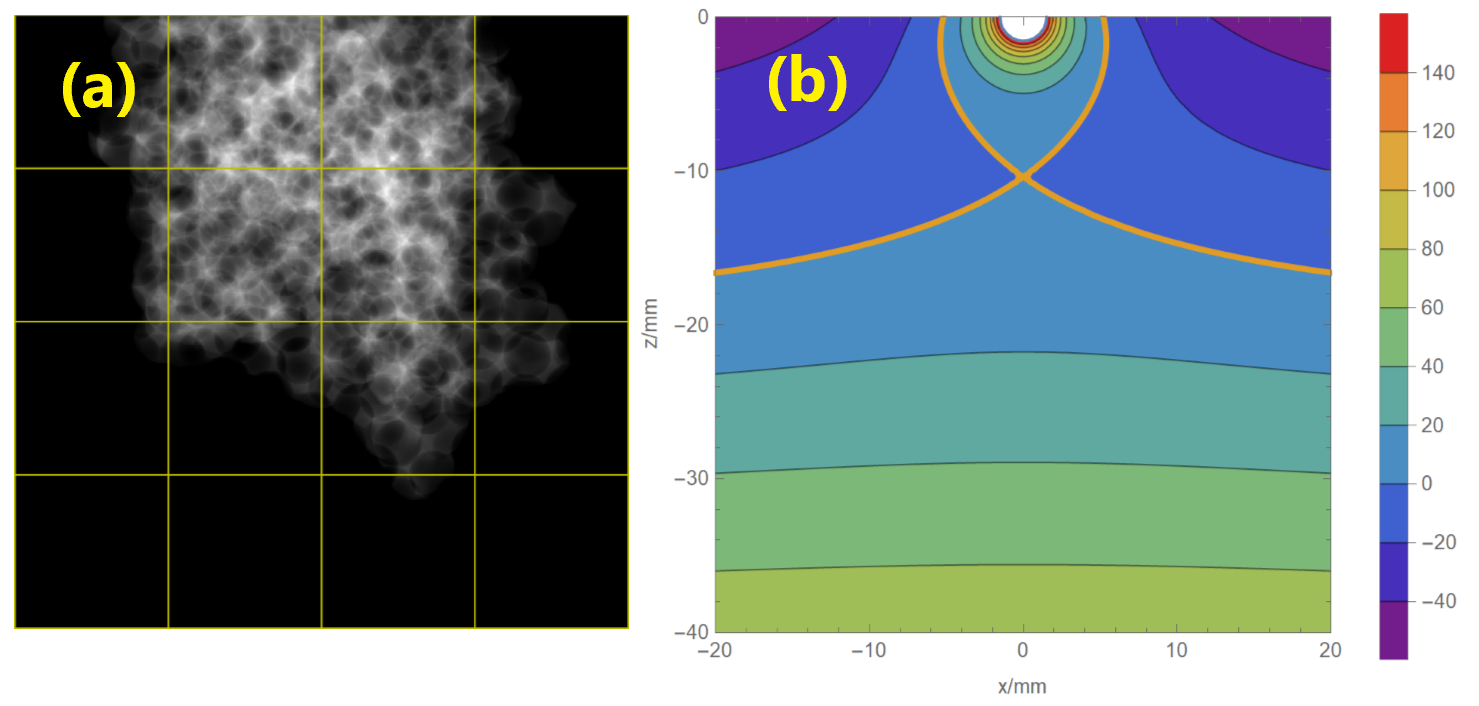}
\end{center}
\caption{The compact center of invading glycerol for $Q=1.0$~ml/min near the time it becomes 
unstable.  Inlet at top center of each part.
{\bf (a):} 
Vertical density projection of the experimental result.
{\bf (b):}  
Contour map of the theoretical capillary-pressure landscape in the vertical plane $(x,0,z)$.
Yellow, crossing curves mark the isobar at the saddle point. Other isobars 
are shown at 20~Pa~$\approx W^{\rm t}$ intervals -- blue: negative, green-red: positive.
The vertical color bar shows the 
pressure differences from the saddle point in Pascal. 
The white semicircle at the top shows the position and diameter of the inlet. 
Both plots are shown on the same scale of 40~mm$\, \times \,$40~mm. 
}\label{fig:theo}
\end{figure}

In Fig.~\ref{fig:theo}, we show an experimental image and a theoretical capillary-pressure 
landscape, both projected onto the vertical $(x,0,z)$ plane. 
They represent a time near that, at which the flow becomes unstable, as in 
Fig.~\ref{fig:3}(b). 
(This is an earlier time than that depicted in Fig.~\ref{fig:blob}.) 
Both images illustrate $Q = 1.0$~ml/min and 
cover the same area of 40~mm $\times$ 40~mm. 
Part (a) is a density projection of the compact center of the 
invading glycerol cluster. (Fingers were removed by an Erosion/Dilation method \cite{HARA87}.) 
Part (b) is a contour map of the capillary-pressure landscape. 
The isobar corresponding to the saddle point  is marked by a thick, yellow 
curve. Other isobars at intervals of 20~Pa~$\approx W^{\rm t}$ are marked by 
color changes. 
The growth is expected to be stable in the positive-pressure region surrounded by the 
closed part of the saddle-point isobar. 
Comparing Figs.~\ref{fig:theo}(a) and (b), the compact glycerol body 
is seen to have a diameter of about 20~mm and to extend no more than about 
$\pm 20$~Pa outside the theoretically expected stable region in the 
capillary-pressure landscape. 
Thus, the agreement between the experimental result and our simple theoretical estimate 
appears satisfactory. 

Further refinements of the experiments and theoretical analysis are in progress.

\section*{Acknowledgments}
We thank G.~M.\ Buend{\'\i}a and A.~J.\ Gurfinkel for useful comments. 
Supported by the Research Council of Norway
through the Center of Excellence funding scheme, Project No.\ 262644.


\end{document}